\documentclass[showpacs,prl,twocolumn]{revtex4}
\usepackage{epsfig,amsmath}
\usepackage{subfigure}
\usepackage{graphicx}
\usepackage{dcolumn}
\usepackage{stmaryrd}
\usepackage{mathrsfs}
\usepackage{pifont}
\usepackage{amsthm}
\usepackage{amssymb}
\usepackage{bm}
\usepackage{latexsym}
\usepackage{hyperref}
\usepackage{color}

\newcommand{\beq}{\begin{equation}}
	\newcommand{\eeq}{\end{equation}}
\newcommand{\beqa}{\begin{eqnarray}}
	\newcommand{\eeqa}{\end{eqnarray}}

\begin{document}

\author{Zhao-Ming Wang$^{1}$\footnote{wangzhaoming@ouc.edu.cn}, Feng-Hua Ren$^{2}$\footnote{renfenghua@qtech.edu.cn}, Mark S. Byrd$^{3}$\footnote{mbyrd@siu.edu}, and Lian-Ao Wu$^{4,5}$\footnote{lianao.wu@ehu.es}}

\affiliation{$^{1}$ College of Physics and Optoelectronic Engineering, Ocean University of China, Qingdao 266100, China \\
$^{2}$ School of Information and Control Engineering, Qingdao University of Technology, Qingdao 266520, China \\
$^{3}$ Department of Physics, Southern Illinois University, Carbondale, Illinois 62901-4401, USA \\
$^{4}$ Department of Physics, University of the Basque Country UPV/EHU, 48080 Bilbao, Spain
IKERBASQUE Basque Foundation for Science, 48013 Bilbao, Spain\\
$^{5}$ EHU Quantum Center, University of the Basque Country UPV/EHU, Leioa, Biscay 48940, Spain}

\title{Hybrid noise protection of logical qubits for universal quantum computation}

\begin{abstract}
Quantum computers now show the promise of surpassing any possible classical machine. However, errors limit this ability and current machines do not have the ability to implement error correcting codes due to the limited number of qubits and limited control. Therefore, dynamical decoupling (DD) and encodings that limit noise with fewer qubits are more promising. For these reasons, we put forth a model of universal quantum computation that has many advantages over strategies that require a large overhead such as the standard quantum error correcting codes. First, we separate collective noise from individual noises on physical qubits and use a decoherence-free subspace (DFS) that uses just two qubits for its encoding to eliminate collective noise. Second, our bath model is very general as it uses a spin-boson type bath but without any Markovian assumption. Third, we are able to either use a steady global magnetic field or to devise a set of DD pulses that remove much of the remaining noise and commute with the logical operations on the encoded qubit. This allows removal of noise while implementing gate operations. Numerical support is given for this hybrid protection strategy which provides an efficient approach to deal with the decoherence problems in quantum computation and is experimentally viable for several current quantum computing systems. This is emphasized by a recent experiment on superconducting qubits which shows promise for increasing the number of gates that can be implemented reliably with some realistic parameter assumptions.  
\end{abstract}

\maketitle

\section{Introduction}
Quantum algorithms \cite{Shor97} have been developed that are much more efficient than the best known classical counterpart. However, it is difficult to perform reliable calculations on a quantum computer due to the well-known problem of decoherence. A variety of schemes for combating the deleterious environment effects have been proposed, including quantum-error correction, DD and DFS/noiseless encodings. Quantum-error correction was shown by Shor \emph{et al}. \cite{Shor95} and subsequent work extended this and demonstrated that techniques exist that can be used to significantly reduce the quantum error rate \cite{Knill, Gottesman, Lidar, Joschka}. However, fully implemented quantum error correction technology cannot be implemented reliably in the noisy intermediate-scale quantum era \cite{NISQ, Sankar} and thus some  decoherence is unavoidable \cite{sc_review_1}. As a result, DD or DFS encodings have been sought to reduce decoherence and extend the functionality of these machines. DD is an active protection strategy and it requires an efficient control of qubits in the presence of noise \cite{Wu2003,Uhrig,Sekiguchi}. On the other hand, DFSs \cite{Duan,Zanardi,Brown} are a type of passive protection. Both of these two schemes have their advantages and limitations, e.g., DFSs encodings may not be easy to design and have been primarily applicable to collective noise. DD schemes can combat both collective and individual noises but the controls may be difficult to implement experimentally \cite{Jing2013,PyshkinSR}. In this paper, we design a hybrid scheme for reliable quantum gate operations. Specifically, we use DD or a steady global magnetic field and DFSs to deal with the $X,Y$ noises and the collective $Z$ noises while enabling reliable computation.  

Computation in subspaces remains unaffected by the interaction with the environment when the interaction Hamiltonian has a certain symmetry \cite{Duan}. Coherence control of two logical qubits encoded in a DFS has been demonstrated, and the DFS encoding has proven to have high fidelity \cite{Henry}. The exchange-only gating scheme of DiVincenzo \emph{et al.} encodes three physical qubits into a logical qubit \cite{DiVincenzo,Kempe}. The number of physical qubits and gate operations will be ameliorated for the XXZ type couplings \cite{Wu,Lidar2002prl,WuA,Nori}. Numerous experimental application of DFSs have been demonstrated, such as in trapped ions \cite{KIELPINSKIV}, in an optical Deutsch-Jozsa algorithm \cite{Mohseni},  in a linear-optical experiment \cite{Kwiat}, and in NMR \cite{Fortunato}.

DFSs can be quite useful for collective baths and recent experiment shows that collective baths do exist.  Charge noises in a superconducting multiqubit circuit chip have been found highly correlated on a length scale over 600 micrometres \cite{Wilen}. Recently, quantum computation in DFS has been proved to be possible and the computation is robust against collective decoherence in quantum systems \cite{Pyshkin}. However, collective baths are special and individual baths \cite{Unruh,JingjunSR} are more common. More generally, individual and collective baths both coexist in many systems. In this paper, we consider a mixed bath, which includes both collective \cite{Duan98,Yavuz} and individual noises \cite{Unruh}. We use a simple encoding scheme with one logical qubit encoded by two physical qubits as in Ref.~\cite{Wu}. Also, the mitigation of noises was addressed by algebraic means
complemented with numerical optimal control in the Markovian regime of Lindblad or Bloch-Redfield type \cite{JPB}. Considering the non-Markovian environmental noises, for the gate operations, we calculate the fidelity dynamics.  On the other hand, both the individual and the collective $X,Y$ noises can be eliminated by a steady global magnetic field on our entire quantum computation, so that we don't require any additional physical operations. Similarly, we can also use the DD technique and a global leakage elimination operator (LEO), formulated as $\sum_i \sigma^{z}_i$  to eliminate both collective and individual $X,Y$ noises. Since this global LEO commutes with all logical operations in the entire quantum computation process, it can be implemented independent of the gating operations. As for the leftover individual $Z$ noise, we study the effects of the environment parameters on the obtainable rotation angle for given fidelity and find the region where gates remain accurate  even if the individual $Z$ noise is relatively strong. We show some threshold within which the universal set of gates works perfectly.

\section{Model}
The total Hamiltonian has the form
\begin{equation}
	H_{tot}=H_{s}+H_{b}+H_{int},
\end{equation}%
where $H_{s}$ is the system Hamiltonian, $H_{b}=\sum_{j=0}^{N}H_{b}^{j}$ is the bath Hamiltonian and $H_{int}$ is the interaction Hamiltonian. Suppose $j=0$ corresponds to a collective bath and $j=1,2,..N$ correspond to $N$-independent individual bath operators, then $H_{b}^{j}=\sum_{k}\omega _{k}^{j}b_{k}^{j\dag}b_{k}^{j}$. $\omega _{k}^{j}$ is the boson's frequency of the $k$th mode and $b_{k}^{j\dag }, b_{k}^{j}$ are the bosonic creation and annihilation operators. The interaction reads
\begin{equation}
	H_{int}=\sum\nolimits_{j,k}(g_{k}^{j\ast }L_{j}^{\dag
	}b_{k}^{j}+g_{k}^{j}L_{j}b_{k}^{j\dag }),
	\label{eq02}
\end{equation}
where $L_{j}$ are system operators, the subscript indicates the coupling to the $j$th bath, and $g^{j}_{k}$ is the coupling constant between the system and $k$th mode of the $j$th bath. Clearly, $L_{0}$ and $g^{0}_{k}$ describe the coupling between the system and the collective bath.

Assume initially that the bath is in a thermal equilibrium state at temperature $
T_{j}$ with the density operator $\rho _{j}(0)=e^{-\beta H_{b}^{j}}/Z_{j},$
where $Z_{j}=$Tr$[e^{-\beta H_{b}^{j}}]$ is the partition function, and $\beta
=1/(K_{B}T_{j})$. The initial density matrix operator is assumed to
be in a product state with the bath, $\rho (0)=\rho _{s}(0)\otimes \rho _{b}(0)=\left\vert \mathbf{\psi }_{0}\right\rangle \left\langle \mathbf{\psi }_{0}\right\vert\bigotimes\limits_{j=1}^{N}\rho _{j}(0)$. Here $\rho _{s}(t)$ and $\rho _{b}(t)$ are the system and bath density
matrix, respectively. Here we assume an uncorrelated initial state between the system and baths. However for a correlated one, the construction of the density matrix does not maintain the positivity of the density matrix \cite{Alicki}. For non-Markovian baths, its asymptotic state strongly depends on the initial conditions \cite{Dariusz}. A measure to quantify the influence of the initial state of an open system on its dynamics is proposed recently, and conditions under which the asymptotic state exists are derived \cite{Wenderoth}. In this paper we use a newly developed theoretical tool, which is referred to as the non-Markovian quantum state diffusion (QSD) approach \cite{Diosi98,YTPRA,Strunz1999,Wang2021,Ren2020}. The non-Markovian master equation is given by \cite{Wang2021}
\begin{eqnarray}
	\frac{\partial }{\partial t}\rho _{s} &=&-i[H_{s},\rho
	_{s}]+\sum\nolimits_{j}\{[L_{j},\rho _{s}\overline{O}_{z}^{j\dag
	}(t)]-[L_{j}^{\dag },\overline{O}_{z}^{j}(t)\rho _{s}]  \notag \\
	&&+[L_{j}^{\dag },\rho _{s}\overline{O}_{w}^{j\dag }(t)]-[L_{j},\overline{O}_{w}^{j}(t)\rho _{s}]\},  \label{eq020}
\end{eqnarray}
where $\overline{O}_{z,(w)}^{j}(t)=\int_{0}^{t}ds\alpha
_{z,(w)}^{j}(t-s)O_{z,(w)}^{j}(t,s)$ and $\alpha _{z,(w)}^{j}(t-s)$ is the
correlation function. The operator $O$ is an \emph{ansatz} and is assumed to be noise-independent here. The operator $O$ is an \emph{ansatz} and is assumed to be noise-independent here. Generally the $\overline{O}$ operators contain noises except for some special cases, such as the case that the system Hamiltonian commutes with the Lindblad operators \cite{YTPRA}. Also, when the bath couples weakly to the system, the noise-dependent  $\overline{O}_{z, (w)}^{j}(t,z^*,w^*)$ operator is approximated well by a time-independent operator $\overline{O}_{z, (w)}^{j}(t)$ \cite{Diosi98,YTPRA,Struntz}. 
Now we use the Lorentz-Drude spectrum as an example to obtain the correlation function, where the spectral density is $J_{j}(\omega )=\frac{\Gamma_{j} }{\pi }\frac{\omega_{j}}{1+(\omega_{j}/\gamma_{j})^{2}}$ \cite{Ritschel,Meier}. Here $\Gamma_{j}$ represents the strength of the $j$th pair system-bath coupling. $\gamma_{j}$ is the characteristic frequency of the $j$th bath. In the high temperature or low frequency limit, references \cite{Wang2021,Ren2020} therefore derives closed equations for $\overline{O}_{z,(w)}^{j}$ to numerically solve the non-Markovian master equation~(\ref{eq020})
\begin{eqnarray}
	\partial \overline{O}_{z}^{j}/\partial t &=&\left(
		\Gamma_{j}T_{j}\gamma_{j}-i\Gamma _{j}\gamma_{j}^{2}\right)L_{j}/2-\gamma_{j}\overline{O}_{z}^{j}  \notag \\
	&&-[iH_{s}+\sum\nolimits_{j}(L_{j}^{\dag }\overline{O}_{z}^{j}+L_{j}%
	\overline{O}_{w}^{j}),\overline{O}_{z}^{j}],  \label{eq027}
\end{eqnarray}
\begin{eqnarray}
	\partial \overline{O}_{w}^{j}/\partial t&=&\Gamma
		_{j}T_{j}\gamma_{j}L_{j}^{\dag }/2-\gamma_{j}\overline{O}%
	_{w}^{j}\notag \\
	&&-[iH_{s}+\sum\nolimits_{j}(L_{j}^{\dag }\overline{O}_{z}^{j}+L_{j}%
	\overline{O}_{w}^{j}),\overline{O}_{w}^{j}].  \label{eq028}
\end{eqnarray}

Simple encodings of one logical qubit into two physical qubits have been suggested to avoid difficult-to-implement single-qubit control terms \cite{Wu}. In this case, the DFS, encoded in one pair of spins, is {$\left\vert 0\right\rangle_L$, $\left\vert 1\right\rangle_L$}. The subscript $L$ denotes the logical qubit. $\left\vert 0\right\rangle_L$=$\left\vert 01\right\rangle$, $\left\vert 1\right\rangle_L$=$\left\vert 10\right\rangle$. The single qubit gates in this subspace can be written in terms of the generators of SU(2) as follows \cite{Wu}: $T_x=(\sigma^x_1\sigma^x_2+\sigma^y_1\sigma^y_2)/2$,  $T_y=(\sigma^y_1\sigma^x_2-\sigma^x_1\sigma^y_2)/2$, $T_z=(\sigma^z_1-\sigma^z_2)/2$.

Now we consider the quantum gate operations of one logical qubit in the presence of noise which is separated into individual and collective parts. The quantum gate fidelity is defined as $F(t)=\int{d\psi(0)\left\langle \psi(0)\right \vert U^{\dagger} \rho_s(t) U \left\vert \psi(0)\right\rangle}$, where $U=e^{-iH_{s}t}\left\vert\psi(0)\right\rangle$ and $\left\vert\psi(0)\right\rangle$ is the arbitrary initial state of the system. $\rho_{s}(t)$ is the system's reduced density matrix in Eq.~(\ref{eq020}).

\section{Results and discussions}
First consider a single-qubit gate with the system Hamiltonian $H_{s}=-JT_{x,(z)}$. Here $J$ is the coupling constant. Suppose both physical spins encounter a collective bath $H_b^{0}$, and each of them couples to an individual bath with Hamiltonian $H_b^{j}$ ($j=1,2$). We use a parameter $\alpha$ to represent the degree of mixing of the two types of baths.  The interaction can then be rewritten as 
\begin{eqnarray}
	H_{int}=\cos^2 \alpha \vec{\sigma_0}\cdot \vec{B_0}+\sin^2 \alpha(\vec{\sigma_1}\cdot \vec{B_1}+\vec{\sigma_2}\cdot \vec{B_2}),
\end{eqnarray}
where the subscript $0$ denotes the collective bath, $1,2$ denote the individual baths of spin $1$ and $2$, respectively. The Pauli vector $\vec{\sigma_i}=(\sigma_i^x,\sigma_i^y,\sigma_i^z)$ and the bath operators $\vec{B_i}=(B_i^x,B_i^y,B_i^z)$ with $B_i^{x(y,z)}$ ($i=0,1,2$) representing the $X(Y,Z)$ noises. Comparing with the interaction in Eq.~(\ref{eq02}), $L_i=\sigma_i$ and $B_i=\sum\nolimits_{k}(g_{k}^{i\ast }b_{k}^{i}+g_{k}^{i}b_{k}^{i\dag})$. Note that the Lindblad operator $L=L^{\dagger}=\sigma_x$ or $\sigma_z$ for spin boson or dephasing. Then the rotating-wave approximation \cite{IntravaiaA,IntravaiaB} is not used here. For generality, we can also use $\alpha_{x,(y,z)}$ to denote the mixture degree of the $X(Y,Z)$ noises. For example, $\alpha_z=0$ ($\pi/2$) correspond to a collective (two individual) $Z$ noise. The bath can be written as $H_{b}=\cos^2 \alpha H_b^0+\sin^2 \alpha (H_b^1+H_b^2)$.

At first, assume there are only $X,Y$ noises present, so 
\begin{eqnarray}
	H_{int}&=&\cos^2 \alpha_x \sigma_0^xB_0^x+\sin^2 \alpha_x\sum\nolimits_{i=1}^{2}\sigma_i^xB_i^x
	\notag \\
	&&+\cos^2 \alpha_y \sigma_0^yB_0^y+\sin^2 \alpha_y\sum\nolimits_{i=1}^{2}\sigma_i^yB_i^y.
\end{eqnarray}  

We observe that if $[\sigma^z_1+\sigma^z_2,T_{x,(y,z)}]=0$, these interactions can be eliminated by adding an LEO Hamiltonian \cite{Wu2002,Wang2020}
\begin{eqnarray}
	H_{LEO}=c(t)(\sigma^z_1+\sigma^z_2),
\end{eqnarray}
where $c(t)$ is the control function. We emphasize that the significant advantage of the LEO Hamiltonian is that, due to $[\sigma^z_1+\sigma^z_2,T_{x,(y,z)}]=0$, adding of such an LEO does not interfere with the gate operation. 

\begin{figure}
	\centering
	\includegraphics[scale=0.14]{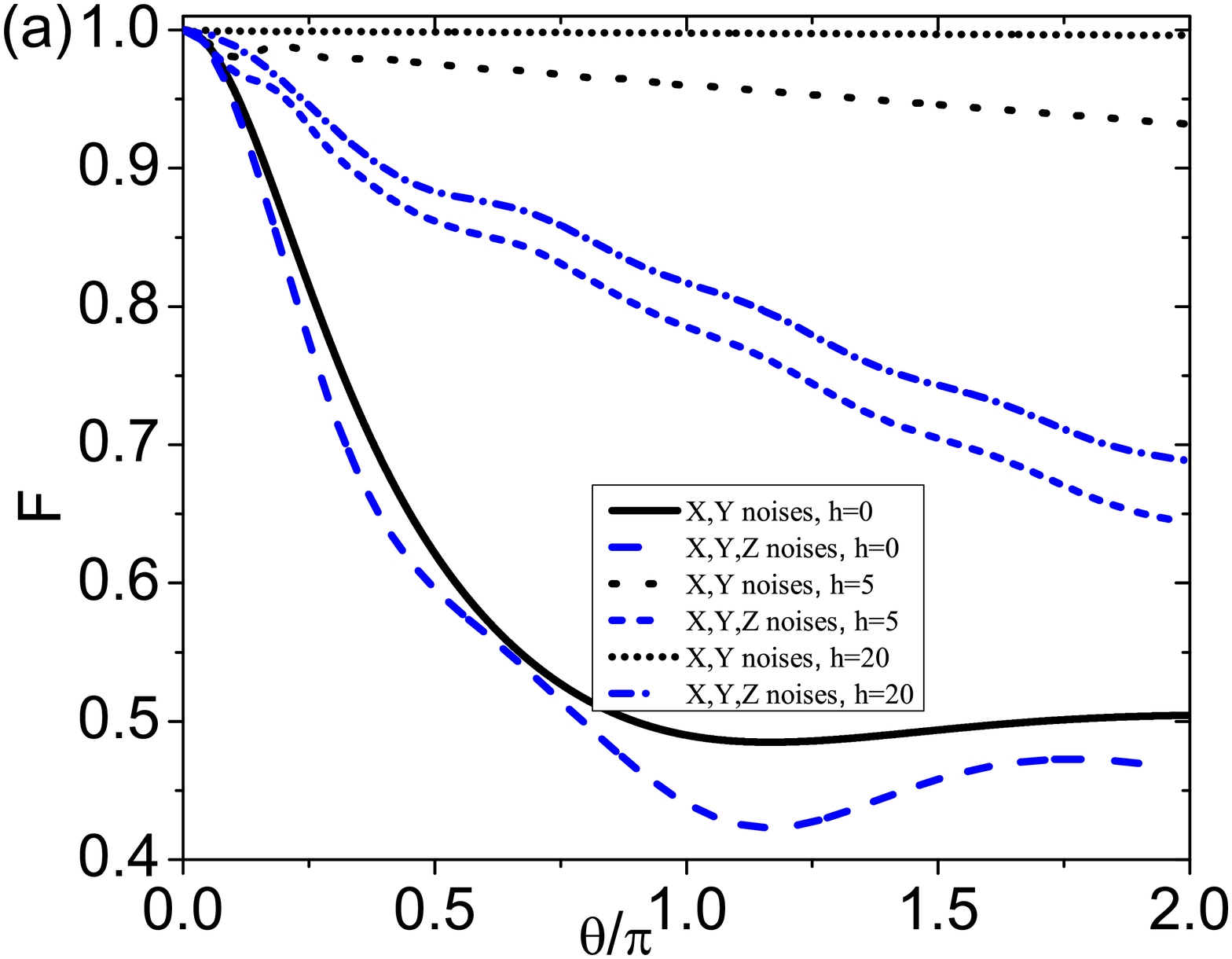}
	\includegraphics[scale=0.14]{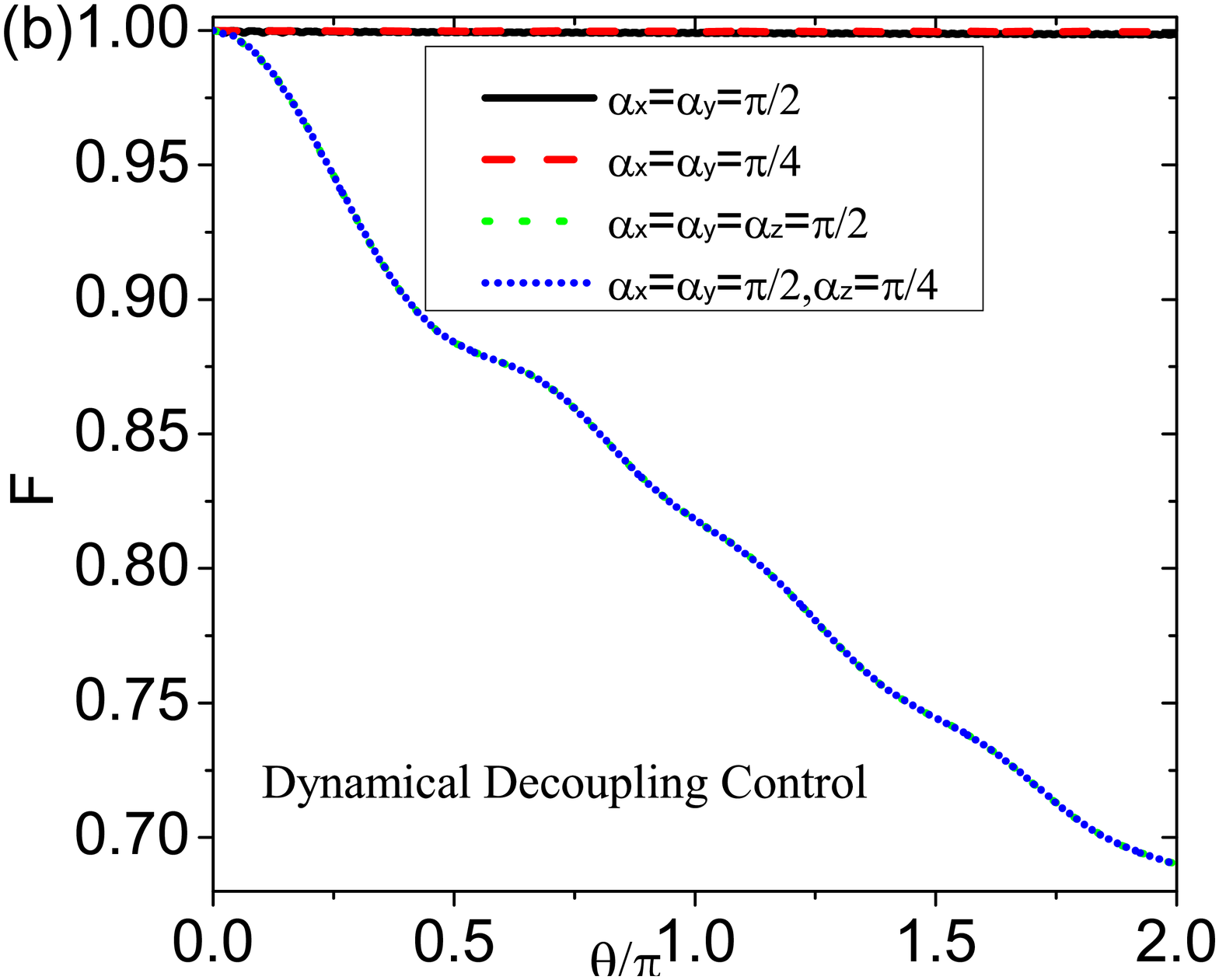}
	\caption{(Color on line) (a) The fidelity $F$ versus $\theta/\pi$ for different parameter $h$ for logical qubits. The $X,Y,Z$ noises are all individual; (b) The fidelity $F$ versus $\theta/\pi$ under DD control. $\theta=Jt$ can be viewed as a rotation angle of the gates. The pulses are present at $\tau$ and absent at the next $\tau$. $H_s=-JT_x$, $\Gamma=0.005$, $\gamma=1$, $T=50$.}
	\label{Fig:1}	
\end{figure}
\begin{figure*}
	\centering
	\includegraphics[scale=0.19]{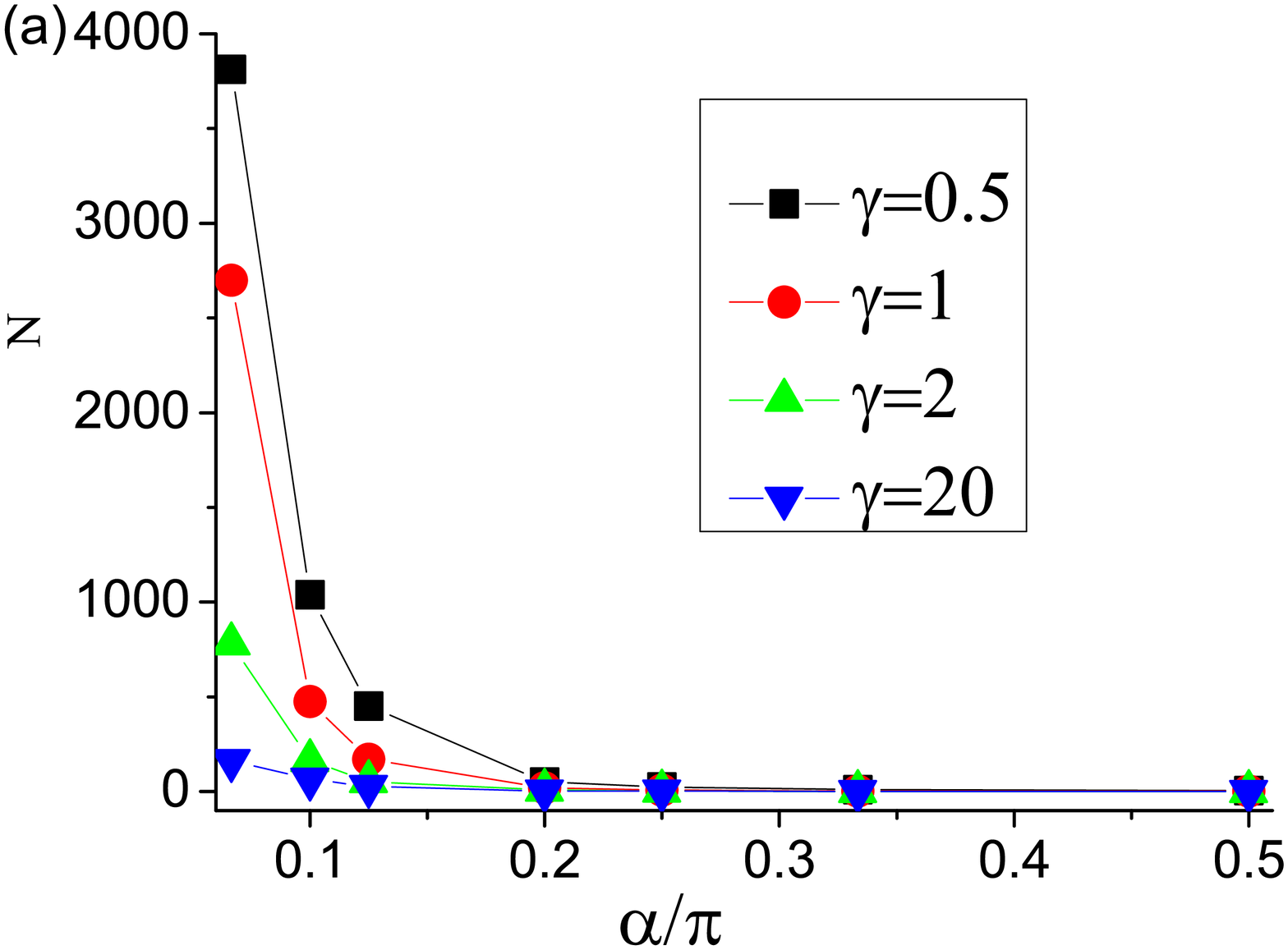}
	\hspace{0.0in}
	\includegraphics[scale=0.19]{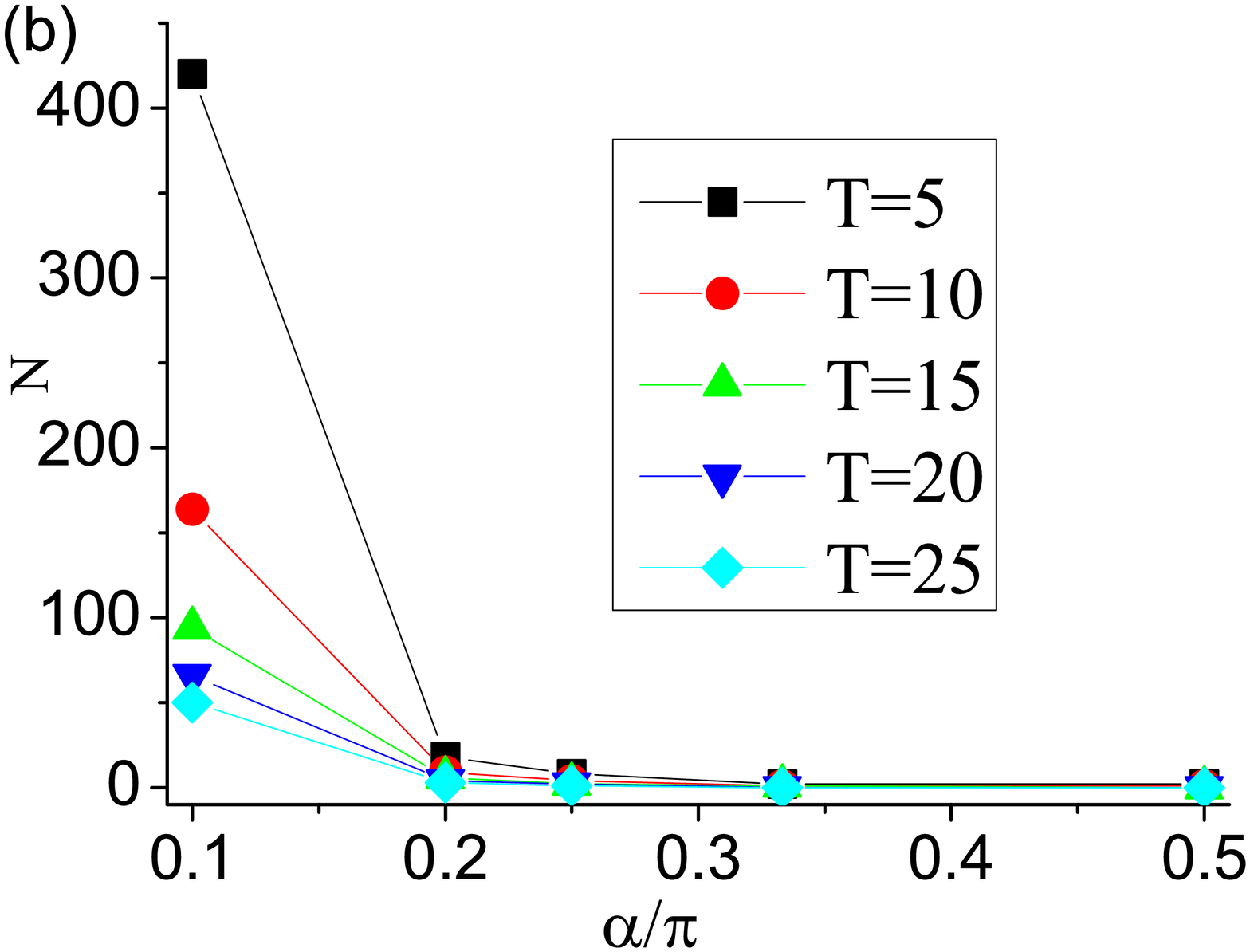}
	\hspace{0.0in}
	\includegraphics[scale=0.19]{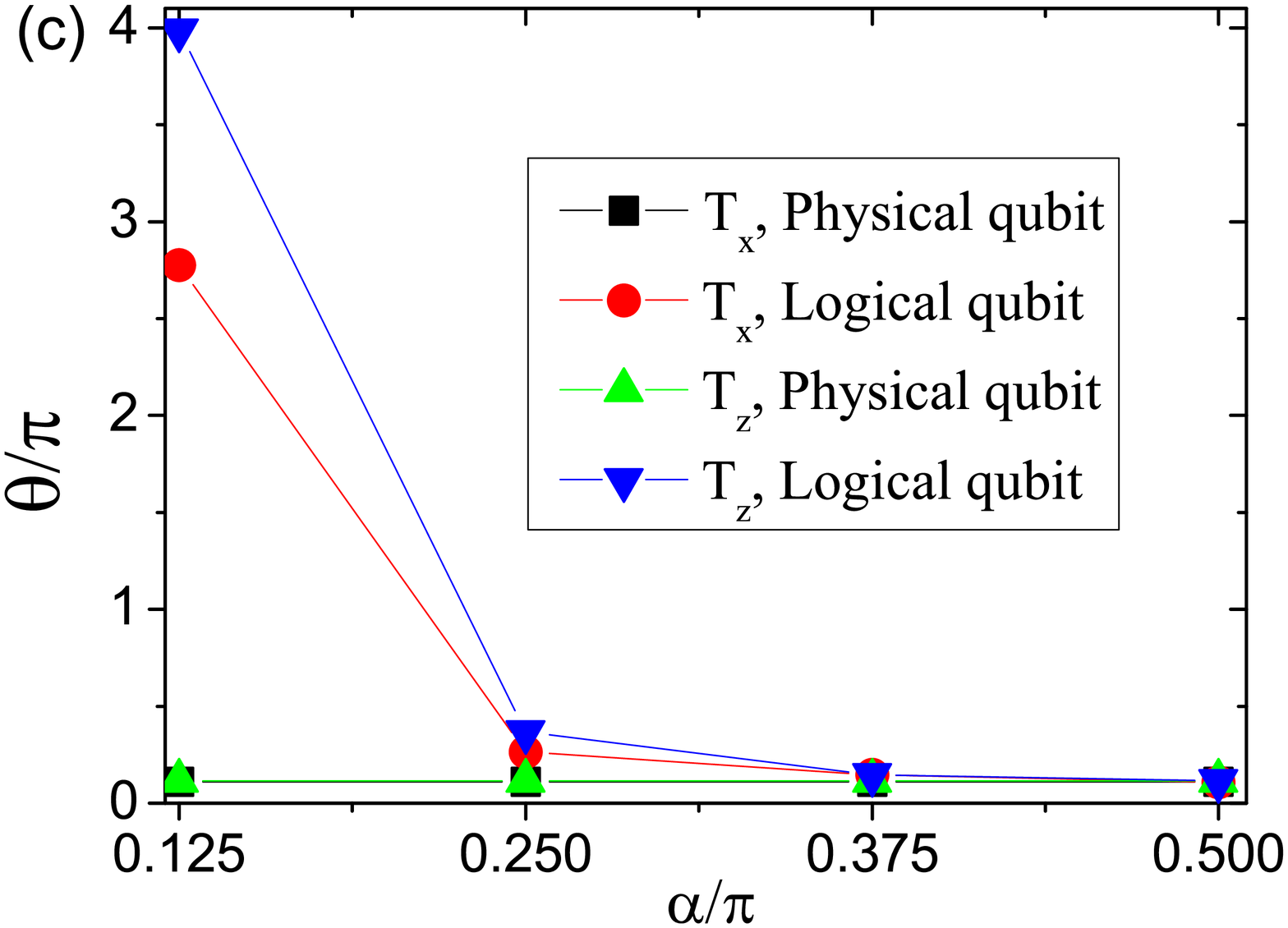}
	\caption{(Color online) The number of gates $N$ ($F=0.95$) versus $\alpha$ with different $\gamma$ (a) or temperature $T$ (b). $T=10$ in (a) and $\gamma=2$ in (b). $\Gamma=0.005$. $H_s=-JT_x$. (c) Comparison of the rotation angle $\theta$ ($F=0.95$) versus $\alpha$ between physical qubit and logical qubit with different $T_{x,(z)}$. $\Gamma=0.005$ ($0.01$) for the logical (physical) qubit. $\gamma=10$, $T=50$, $L=\sigma^z$.}
	\label{Fig:2}
\end{figure*}

Suppose the control function is a constant $c(t)=h$, here $h$, for example, could be the magnetic field. In Fig.~\ref{Fig:1} (a), we plot the fidelity $F$ as a function of the normalized time $\theta/\pi$ for different $h$. Here $\theta=Jt$ can be viewed as a rotation angle and is taken to be $\theta=2\pi$ in the total evolution. We assume the collective bath has the same magnitude as the individual baths, i.e., $\Gamma_{0}=\Gamma_{1}=\Gamma_{2}=\Gamma$, $\gamma_{0}=\gamma_{1}=\gamma_{2}=\gamma$, and $T_{0}=T_{1}=T_{2}=T$. The environmental parameters are taken to be $\Gamma=0.005$, $\gamma=1$, $T=50$. If there is no control, the fidelity will decrease quickly with increasing $\theta$ for $H_s=-JT_x$. As expected, the $X,Y$ or $X,Y,Z$ noises will significantly reduce the gate fidelity. 

We next compare this with control added. For simplicity, we only add a constant pulse. Noting $h >> J$ and that this control does not affect the gate operation. (They commute.)  Fig.~\ref{Fig:1}(a) shows that with the increasing $h$, $F$ has a drastic increase. When $h=20$, $F=1$ for $X,Y$ noises, indicating the interaction has been removed. When it also has $Z$ noise, the control is not as effective. We also check the case $H_s=-JT_z$ and find similar behavior to that in Fig.~\ref{Fig:1}(a). The constant pulse in the above discussion is often difficult to approximate well in experiments. Bang-Bang are idealistic since they assume a relatively strong, fast pulse sequences $c(t)=\pi\sum_{i}\delta(t-\tau_i)/2$ \cite{PyshkinSR}. (They must be strong relative to the natural, or drift, Hamiltonian.) Thus these are also often difficult to be implement experimentally \cite{Jing2013}. Nonperturbative DD which uses a finite pulse intensity and finite pulse intervals is much more practical for effective control \cite{Jing2013, Wang2012}. Next we show how nonperturbative DD can be used to eliminate $X,Y$ noises.  

In the numerical simulation, we use rectangular pulses and the above LEO Hamiltonian to simulate a $\delta$-function pulse. We use $c(t)=50$ (even $n$) and $c(t)=0$ for $n\tau<t<(n+1)\tau$ (odd $n$), $\tau=0.01 \pi$. For this choice, the integral satisfies $\int_{0}^{\tau}c(s)ds=\pi/2$ which is required by theory in one control period $\tau$ \cite{Wang2020}. However, there is no need to stick to $\pi/2$, nonperturbative DD only requires a large constant \cite{Jing2013, Wang2012}, the control function can even be noisy \cite{Jing2014, Long}. In Fig.~\ref{Fig:1}(b), we plot the fidelity $F$ versus the parameter $\theta/\pi$ under DD control. The results again show that the simulated DD pulses are effective to remove the $X,Y$ noises ($\alpha_x=\alpha_y=\pi/2$ or $\alpha_x=\alpha_y=\pi/4$ in Fig.~\ref{Fig:1}(b)), both for individual and collective types.  As expected, it fails for the case that the $X,Y,Z$ noises ($\alpha_x=\alpha_y=\alpha_z=\pi/2$) in Fig.~\ref{Fig:1}(b)). We also plot the case where we have both the individual $X,Y,Z$ noises plus collective $Z$ noise ($\alpha_x=\alpha_y=\pi/2,\alpha_z=\pi/4$). We find that in this case the protection of the DFS encodings is still effective and the fidelity evolution for these two cases are the same. \emph{To summarize, the advantage of our hybrid strategy is that if there are $X,Y$ noises and only collective $Z$ noise, reliable quantum gate operation can be realized by adding a control that effectively removes the interaction and the effects of the noises can be completely eliminated.}

From the above analysis, the individual $Z$ noise is not easily eliminated.  It is therefore important to check how the mixture of the individual and collective $Z$ noises affect the number of the gates $N$ that can be implemented for certain threshold fidelity, e.g., $F(\theta)=0.95$. Recently, a quantum error correction threshold of $4.7\%$ using a clustering decoder has been found for a depolarizing noise model \cite{Alexis}. In the following part we will only discuss $Z$ noise. Now for collective bath $\sigma_0=\sigma^z_1+\sigma^z_2$, and for individual baths $\sigma_1=\sigma^z_1, \sigma_2=\sigma^z_2$. In Figs.~\ref{Fig:2}(a) and (b), we plot the number of the gates $N$ versus $\alpha$ for different $\gamma$ and $T$, respectively. We take $H_s=-JT_x$ as an example, and for $H_s=-JT_z$ we obtain similar results. $N$ decreases with increasing parameter $\alpha$, which shows that $N$ can be dramatically enhanced by increasing the collective bath ratio $\alpha$. We emphasize that an increase in this ratio has been realized in recent work \cite{Wilen}. Our results support that thousands of gates can be implemented for a small $\alpha$. From Figs.~\ref{Fig:2}(a), non-Markovinity of the baths play an important role in boosting the achievable number of gates $N$. Fig.~\ref{Fig:2}(b) shows that the $N$ decreases with increasing temperature $T$ as expected.

DFS encodings provide passive protection for collective noises. Then to what degree does the noise differ from that for the physical qubit?  For a single physical qubit gate, the corresponding Hamiltonian is $H=J\sigma^{x,(z)}+\sigma^zB+H_B$. In Fig.~\ref{Fig:2}(c), we compare the achievable $\theta$ for the physical qubit and the logical qubit. $\theta$ versus $\alpha$ for $T_x$ or $T_z$ is plotted. $\gamma=10$, $T=50$ and $J=1$ are used for both cases, while we take $\Gamma=0.005$ (1.0) for the logical (physical) qubit. For $T_{x,z}$, there is only an individual bath for one single physical qubit. Then $\theta$ is a constant. However, for the logical qubit, it has both collective and individual baths. For $T_x$, $\alpha=\pi/8$, and $\theta/\pi=4$ and for $T_z$, $\theta/\pi=2.6$. Increasing $\alpha$, $\theta$ begins to decrease. When $\alpha=\pi/2$, i.e., there are only individual baths for the logical qubit, we find that the dynamics are the same for the two cases: the interaction strength parameter $2\Gamma$ (physical bit) as $\Gamma$ (logical qubit).

\begin{figure} 
	
	\centerline{\includegraphics[width=0.70\columnwidth]{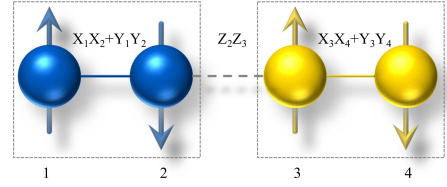}}
	
	\caption{(Color online) Two logical qubits which consist of two pairs of spins. The solid line in the spin pair represents the $XX+YY$ coupling and the dashed line between the two pairs represents the $ZZ$ coupling.}
	\label{Fig:3}
\end{figure}

\begin{figure}
	\centering
	\includegraphics[scale=0.145]{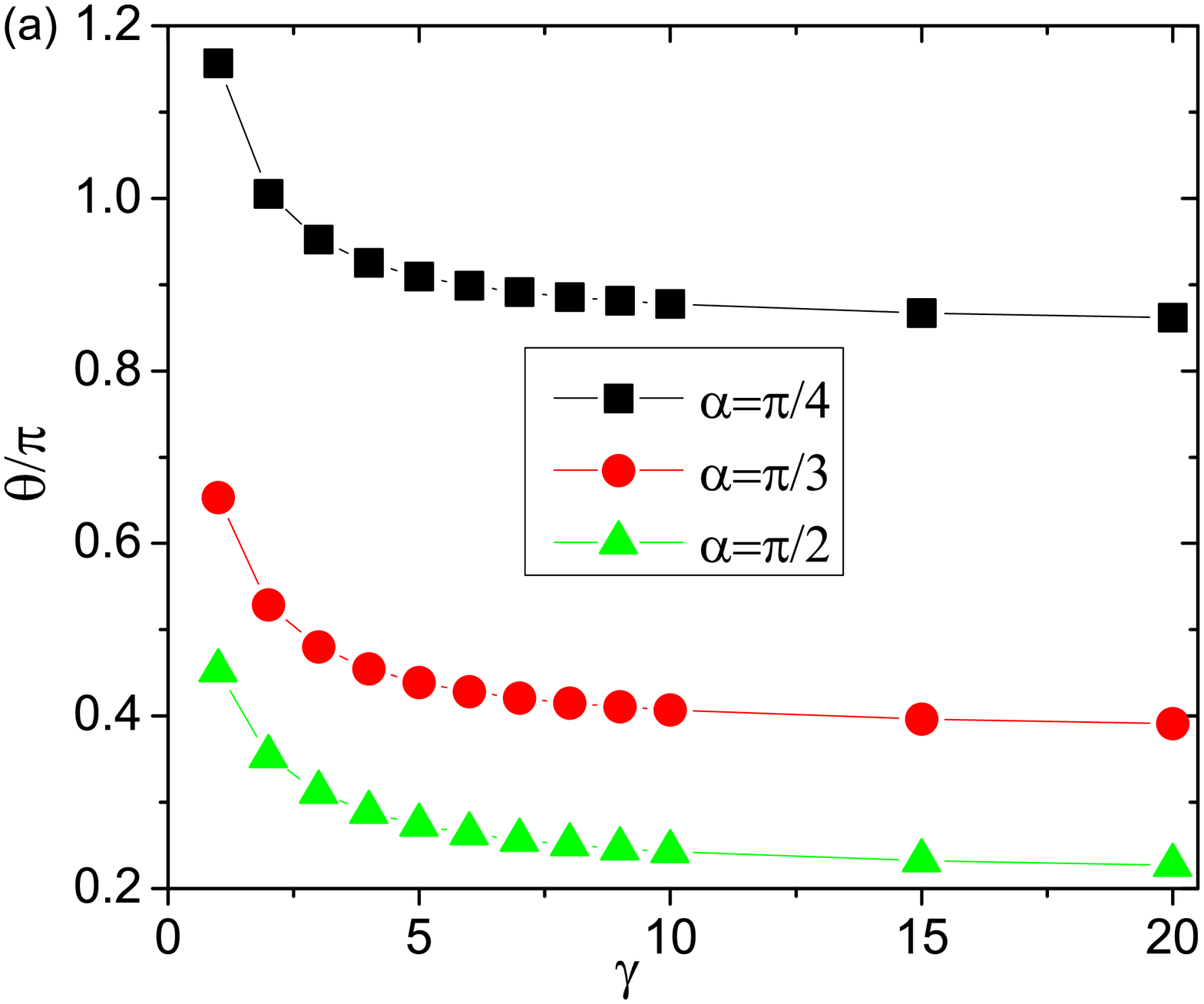}
	\hspace{0.000in}
	\includegraphics[scale=0.145]{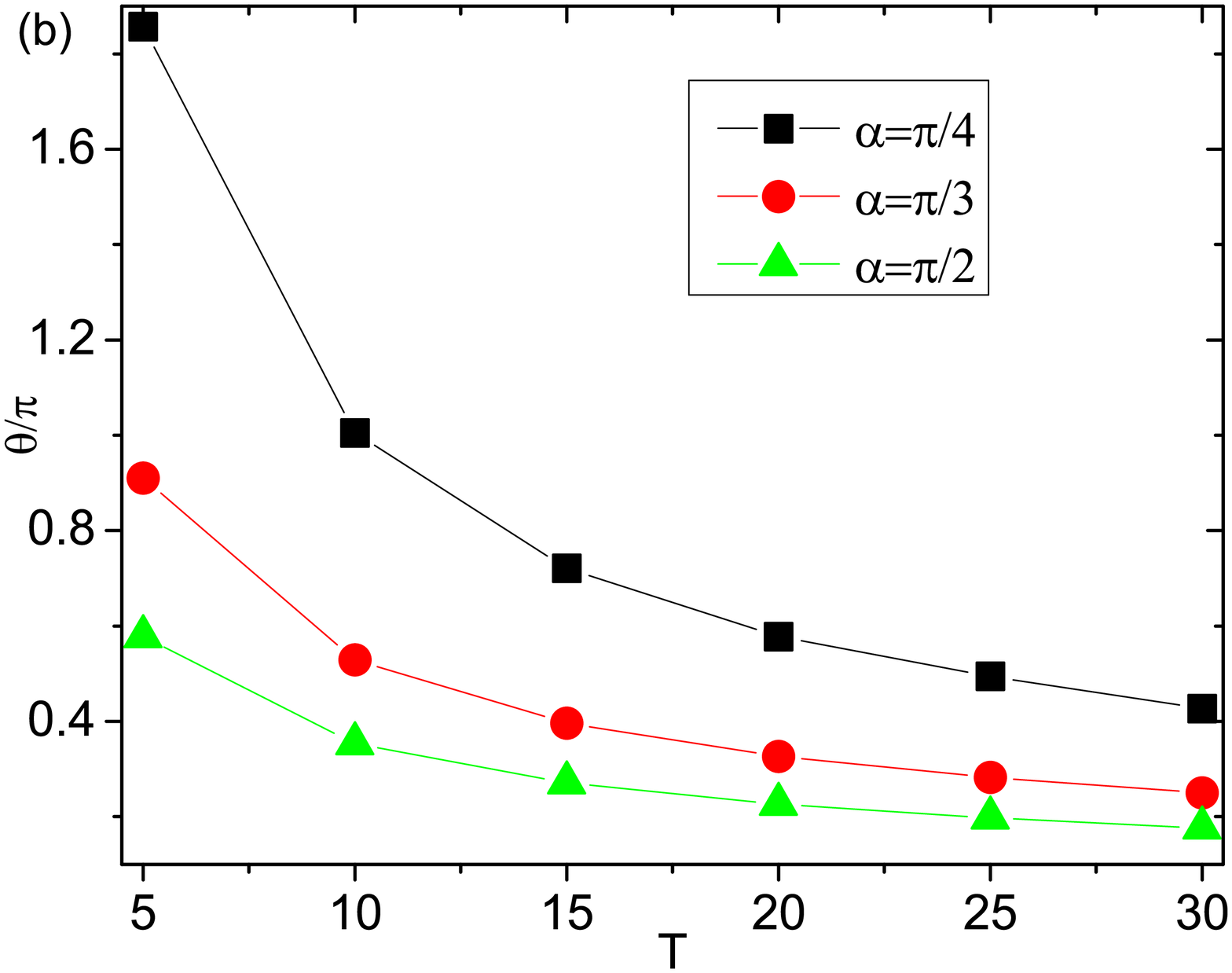}
	\caption{(Color online) The rotation angle $\theta$ ($F=0.95$) versus $\gamma$ (a) and temperature $T$ (b) with different $\alpha$ values. $T=10$, $\Gamma=0.005$ in (a) and $\gamma=2$, $\Gamma=0.005$ in (b). $H_s=-JT_{z1}T_{z2}$.} 
	\label{Fig:4}
\end{figure}

Now let us consider two pairs of spins, each pair encodes one logical qubit. Assume spins 1 and 2 (3 and 4) belong to the first (second) logical qubit (See Fig.~\ref{Fig:3}). Controlled operations between two logical qubit can be made by $T_{z1}T_{z2}=-Z_{2}Z_3$, which implements a two-qubit entangling gate. The operator $e^{i\theta T_{z1}T_{z2}}$ gives a controlled phase gate when $\theta=\pi/2$. Together with a Hadamard gate, we can get a CNOT \cite{Pyshkin}. We point out that the couplings between the spins can be tuned by adding an external field \cite{Jepsen}.  For the two logical qubits, the system Hamiltonian is $H_s=-JT_{z1}T_{z2}$. In Figs.~\ref{Fig:4}(a) and (b) we plot the angle $\theta/\pi$ versus $\gamma$ and $T$ for the two-qubit gate. Fig.~\ref{Fig:4} shows that the obtainable angle $\theta$ decreases with increasing $\gamma$ and $T$. For certain parameters, $\theta$ also decreases with increasing $\alpha$. The parameter dependence shows a behavior similar to the single-qubit gate. 

\section{Experimental parameters}
The dimensionless parameters used in this paper can be converted into a dimensional form for a comparison with a recent experiment using superconducting qubits \cite{Yan19}. For the  Hamiltonian of a single-qubit gate, each qubit can be regarded as a spin-1/2 system, in this case, the coupling $J$ is the nearest-neighbor hopping strength. The typical coupling is around $12.5$ MHz \cite{Yan19} and $\hbar/J \simeq 0.08$ $\mu s$. The tunable $z$-axis coupling between qubits can be performed with an additional intermediate qubit mode and the coupling strength is governed by the flux bias applied to the coupler \cite{chen2014qubit, huang2020superconducting}. The strength $J_z$ can be tuned to $10$ MHz in a high-coherence superconducting circuit \cite{Kounalakis}, which is near the coupling strength $J$. The $z$-axis rotations on individual qubits can be performed by modulating the microwave and local magnetic fields \cite{sc_review_1}.  Increasing the collective bath ratio in the experiment \cite{Wilen} can greatly enhance the number of gates that can be applied, enhancing the ability to perform computations.

\section{Conclusions}
We have designed a hybrid error reduction method using very low overhead.  Motivated by recent experiments, the method uses passive protection (a DFS) to reduce collective errors while employing active correction (LEOs) for individual noise. The calculation shows that high fidelity can be obtained for low temperature, and high non-Markovianity of the baths.  This strategy shows promise theoretically, and numerically.  We have also provided evidence that suggests significant improvement for experiments involving superconducting qubits.  

\section{ACKNOWLEDGMENTS}
This work was supported by the Natural Science Foundation of Shandong Province (Grant No. ZR2021LLZ004) and Fundamental Research Funds for the Central Universities
(Grant No. 202364008)., the grant PID2021-126273NB-I00 funded by MCIN/AEI/10.13039/501100011033, and by “ERDF A way of making Europe” and the Basque Government through grant number IT1470-22.

\end{document}